\documentclass[prb, twocolumn]{revtex4}%
\usepackage{times}
\usepackage{graphicx}
\usepackage{dcolumn}
\usepackage{bm}
\usepackage[usenames,dvipsnames]{color}
\usepackage{amsmath}
\usepackage{amsfonts}
\usepackage{amsthm}
\usepackage{amssymb}
\usepackage{mathrsfs}
\usepackage{subfigure}
\usepackage{ulem}
\usepackage{verbatim}

\def\bra#1{\mathinner{\langle{#1}|}}
\def\ket#1{\mathinner{|{#1}\rangle}}

\def\bra#1{\mathinner{\langle{#1}|}}
\def\ket#1{\mathinner{|{#1}\rangle}}

\begin{document}

\title{Influence of errors on the transport of quantum information through distant quantum dot spin qubits}
\author{Iann Cunha}
\author{Leonardo Kleber Castelano}
\email{lkcastelano@ufscar.br}
\affiliation{Departamento de F\'{\i}sica, Universidade Federal de S\~ao Carlos, 13565-905, S\~ao Carlos, SP, Brazil}

\date{\today}
\begin{abstract}
The ability to connect distant qubits plays a fundamental role in quantum computing. Therefore, quantum systems candidates for quantum computation must be able to interact all their constituent qubits. Here, we model the quantum dot spin qubits by a spin chain with nearest-neighbors interaction. Within this model, we can perform the interaction of distant qubits by the action of consecutive SWAP gates. The SWAP gate exchange the information of two different qubits and it is obtained by a time-dependent interaction of nearest-neighbors qubits that is switched on and off as the quantum information is propagated through the system. By using this scheme, we also are able to implement the CNOT gate, which is a fundamental gate to obtain universal quantum computation. These gates are probed in a system free from decoherence, which provides a very efficient connection between distant qubits. Furthermore, we analyze the situation when the dissipation is present. To perform such a task, we consider dephasing and amplitude-damping types of errors in each site of the spin chain. We found that the order of the SWAP and CNOT gates is important and it can lead to a relevant difference in fidelity when the number of qubits is large. 
\end{abstract}
\maketitle
\section{Introduction}
Silicon quantum dot (QD) semiconductor spin qubits are a promising candidate for a quantum computer's architecture~\cite{PhysRevA.57.120} because of the advantages of long coherence times and the possibility of scaling found in this platform. Si-based systems also have the advantage of been easily integrated into the current semiconductor industry. One of the leading difficulties related to this platform for universal quantum computing is the ability to connect distant spin qubits. A potential solution to overcome this challenge is the quantum network, where the connectivity of qubits is achieved either by charge interaction~\cite{PhysRevB.102.195418,Sigillito2019} or by interfacing nodes via optical links~\cite{PhysRevA.89.022317,PhysRevX.4.041041}. When dealing with the connectivity employing the charge interaction, the wave function of electrons localized in neighbor sites must overlap, which leads to only nearest-neighbor qubits interaction. This short-range spatial interaction restricts the connection of long-distance qubits and represents a significant barrier to implementing error correction~\cite{RevModPhys.87.307,PhysRevLett.77.2585}, quantum random access memory~\cite{PhysRevLett.100.160501}, and some quantum algorithms~\cite{Montanaro2016,Bauer2020}.
Nonetheless, the quantum information can be shuttled across the device through SWAP gates, that expands the connection between distant qubits~\cite{Sigillito2019}.

In this paper, we model the QD qubits by a spin chain, where the charge interaction between neighbors is described by a Heinsenberg Hamiltonian. Moreover, the flow of information within the spin chain is carried by the SWAP gate, which is achieved by switching on and off the interaction between nearest-neighbors. Thus, the interaction between qubits is controlled by a time-dependent Gaussian pulse, whose parameters are optimized to implement the quantum gates. We are using a specific type of pulse, but this approach can be useful in real systems if optimal control theory is employed to engineer the pulses~\cite{PhysRevB.97.235301,PhysRevA.104.062415}.
When the system is free from noise, the fidelity related to the transport of quantum information in the spin chain is very robust. Moreover, we include a dephasing or an amplitude-damping channel in each site of the spin chain to describe the decoherence throughout the quantum network. Very recently, a non-Hermitian Hamiltonian was adopted to probe effects of dissipation and gate timing
errors on the fidelity of a sequence of SWAP gates on a chain of interacting qubits~\cite{PhysRevB.105.155411}. In this paper, we consider the SWAP and the CNOT gates, which are very important two-qubit gates because the SWAP transport the information along the chain and the CNOT is fundamental to achieve universal quantum computing~\cite{NielsenChuang}. Moreover, we investigate the role of exchanging the order of CNOT and SWAP gates when noise is taken into account. We find that the order of the quantum gates application is important and can result in a relevant difference of fidelity when the number of qubits is large.

\section{Theoretical Model}
\subsection{Heisenberg spin chain model}

We map the QD spin qubits in a spin chain containing N spin-$1/2
$ particles coupled to each other by nearest-neighbors interactions, which are accordingly tuned through pulses that implement quantum gates. Therefore, we have a time-dependent Hamiltonian
\begin{equation}
    \mathcal{H}(t) = \sum_{n=1}^{N-1} H_{n, n+1}(t),
    \label{eq:general_hamiltonian}
\end{equation}
where the interaction between the $n$-th and the $(n+1)$-th particles is provided by
\begin{equation}
    \text{H}_{n, n+1}(t) = \sum_{i,j} J_{n}^{i, j}(t)  \sigma_{n}^{i} \sigma_{n+1}^{j},\label{interaction}
\end{equation}
where
\begin{equation}
\begin{split}
    \sigma_{n}^{i}\sigma_{n+1}^{j} = 
    \mathbf{I}^{\otimes n-1} \otimes \sigma^{i} \otimes \sigma^{j} \otimes \mathbf{I}^{\otimes N-n-1},
\end{split}
\end{equation}
and $\sigma^{i}$ represents the respective Pauli operator with $i,j\in\{0,x, y, z\}$.
Also, we intend to analyse the influence of noise on the performance of the quantum gates. The noise is included via the Lindblad quantum master equation
\begin{equation}
    \frac{\partial \rho}{\partial t} = -i [\mathcal{H}, \rho] + \mathcal{D}(\rho),
    \label{eq:lindblad_master_equation}
\end{equation}
where $\rho$ is the density matrix and $\mathcal{D}$ is the dissipator operator given by
\begin{equation}
    \mathcal{D}(\rho) = \sum_{n=1}^{N} \left( L_{n}\rho L_{n}^{\dag} - \frac{1}{2} \{ L_{n}^{\dag} L_{n}, \rho \} \right).\label{eq:dissipator}
\end{equation}
We consider the qubits subjected to the dephasing and the amplitude damping types of noise, where the Lindblad operators are $L_n = \sqrt{\gamma} \sigma_{n}^{z}$ and $L_n = \sqrt{\gamma} \sigma_{n}^{-}$, respectively. The decay rate $\gamma$ is assumed to be constant for all qubits and $\sigma_{n}^{k}= \mathbf{I}^{\otimes n-1} \otimes \sigma^{k}\otimes \mathbf{I}^{\otimes N-n}$ for k=$z,-$. In the amplitude damping noise the lowering operator $\sigma_-=(\sigma_x-i\sigma_y)/2$ describes the process of spontaneous emission.

\subsection{Implementation of SWAP and CNOT gates}
In this section, we discuss two important gates related to quantum computing, the SWAP and CNOT gates. 
The SWAP gate swaps the information between two qubits and it is very useful to transport quantum information around the spin chain. On the other hand, the CNOT gate is fundamental to implement universal quantum computing. 
To implement these gates, we model the time-dependent coupling between nearest-neighbors in eq.~(\ref{interaction}) by Gaussian pulses. Because these gates act only on two qubits, it is enough to search for pulses such as
\begin{equation}
    J_{G}^{i,j}(t) = A_{G}^{i,j} \exp \left( -\frac{(t-\tau)^2}{W_{G}^{i,j}} \right),
\end{equation}
where $A_{G}^{i,j}$ is the peak amplitude, $W_{G}^{i,j}$ is the width, $\tau$ is the center position of the pulse, and G=SWAP or CNOT.
Our first goal is to find $\mathcal{H}(t)$ for two qubits, such that the initial input state $\ket{\psi_{in}}$ evolves to the final output state $\ket{\psi_{out}}$, given by $\ket{\psi_{out}} = U(t_0, T) \ket{\psi_{in}}$, where $U(t_0, T)$ is the unitary time evolution operator and $t_0$ ($T$) is the initial (final) evolution time. Moreover, the final output state must corresponds to a desired state $\ket{\psi_{d}} = G \ket{\psi_{in}}$, where $G$ represents the quantum gate. To probe the equivalence between the evolved state $\rho_{out}=\ket{\psi_{out}}\bra{\psi_{out}}$ and the desired state $\rho_{d}=\ket{\psi_{d}}\bra{\psi_{d}}$, we employ the fidelity
\begin{equation}
    F(\rho_{out}, \rho_{d}) = \left( \text{Tr} \sqrt{\sqrt{\rho_{out}} \rho_{d} \sqrt{\rho_{out}}} \right)^2.\label{eq:fid}
\end{equation}
For a two qubit system with an initial state as $|\psi_{in}\rangle = | q_1 \rangle | q_2 \rangle$, the SWAP gate acts in the following way
\begin{equation}
    \text{SWAP} \left( | q_1 \rangle | q_2 \rangle \right) = | q_2 \rangle | q_1 \rangle,
\end{equation}
where $| q_1 \rangle$ and $| q_2 \rangle$ are arbitrary states
For two qubits, we found by inspection that the Hamiltonian able to implement such a gate is
\begin{equation}
    H_{\text{SWAP}}(t) = J_{\text{SWAP}}(t) \left( \sigma^x \sigma^x +  \sigma^y \sigma^y + \sigma^z \sigma^z \right),
    \label{eq:Hswap}
\end{equation}
with the time-dependent pulse
\begin{equation}
    J^{\text{SWAP}}(t) = A_{\text{SWAP}} \exp \left( -\frac{(t-\tau)^2}{W_{\text{SWAP}}} \right).
    \label{eq:Jswap}
\end{equation}

Similarly, the CNOT gate for the initial state $\ket{\psi_{in}} = \ket{q_1} \ket{q_2}$ acts as
\begin{equation}
    \text{CNOT} \left( \ket{q_1} \ket{q_2} \right) = \ket{q_1}\ket{\tilde{q}_2},
\end{equation}
where $\ket{q_1}$ is the control qubit and $\ket{q_2}$ is the target qubit. The CNOT gate flips the target qubit if and only if the control qubit is $\ket{1}$.
By inspection, we find that the Hamiltonian able to implement the CNOT gate has the following form
\begin{equation}
\begin{split}
    H_{\text{CNOT}}(t) &= J^{1}_{\text{CNOT}}(t) \left( \mathbf{I} \sigma^x + \sigma^z \mathbf{I} \right) \\
    &+ J^{2}_{\text{CNOT}}(t) \left( \sigma^z \sigma^x \right),
\end{split}
\label{eq:Hcnot}
\end{equation}
where,
\begin{equation}
    J^{k}_{\text{CNOT}}(t) = A^{k}_{\text{CNOT}} \exp \left( - \frac{(t - \tau)^2}{W^{k}_\text{CNOT}} \right)\; k=1,2.\label{eq:Jcnot}
\end{equation}
The CNOT Hamiltonian in eq.~(\ref{eq:Hcnot}) can be written in a more familiar form, by rotating the first qubit in $\pi/2$ around the $y-$direction. Thus, the rotated initial state becomes $| \widetilde{\psi}_{in} \rangle = R_{y}| q_1 \rangle | q_2 \rangle = \left( R_{y} \otimes \mathbf{I} \right) | \psi_{in} \rangle$ and the rotated Hamiltonian is 
\begin{equation}
\begin{split}
    \widetilde{H}_{\text{CNOT}} &= J^{1}_{\text{CNOT}}(t) \left( \mathbf{I} \sigma^x - \sigma^x \mathbf{I} \right) \\
    &J^{2}_{\text{CNOT}}(t)  \sigma^x \sigma^x,
\end{split}
\label{eq:Hcnot-unmix}
\end{equation}
 where $R_{y} = e^{i\frac{\pi}{2}\sigma_y}$.
 The terms of both Hamiltonians that appear in eqs.(\ref{eq:Hswap}) and (\ref{eq:Hcnot}) were found by inspection. The structure of the SWAP Hamiltonian is easy to guess, but the CNOT Hamiltonian is more complicated and it was found after many trial and error. In ref.~\cite{Makhlin2002} there is another kind of structure for the CNOT gate, but we could not implement such a form by using the optimization of the Gaussian pulses parameters.
\section{Numerical Results}
To numerically solve the time evolution equation~(\ref{eq:lindblad_master_equation}), we first need to define some parameters. We adopt the time scale $\tau_0$, which coincides to the one pulse duration time, if not mentioned otherwise. In this case, the center position of the pulse is $\tau=\tau_0/2$ in eqs.~(\ref{eq:Jswap}) and(\ref{eq:Jcnot}). All other variables can be written in terms of $\omega_0=1/\tau_0$ and/or $\hbar$. Both SWAP and CNOT gates for a two qubit system were found by means of numerical optimization of the parameters related to the Gaussian pulses and the time evolution operator considering the unitary evolution, {\it i.e.}, removing the dissipator term in eq.~(\ref{eq:lindblad_master_equation}). We search for parameters that maximize the fidelity (eq.~(\ref{eq:fid})) between the desired state resulting from the action of the gate $G=$SWAP or CNOT and the time evolved state considering the set of initial states $\left\{|\psi_1\rangle=|\downarrow\downarrow\rangle,|\psi_2\rangle=|\downarrow\uparrow\rangle,|\psi_3\rangle=|\uparrow\downarrow\rangle,|\psi_4\rangle=|\uparrow\uparrow\rangle,|\psi_5\rangle\right\}$, where $|\psi_5\rangle=(|\psi_1\rangle+|\psi_2\rangle+|\psi_3\rangle+|\psi_4\rangle)/2$, which accounts for avoiding undesirable phase errors~\cite{PhysRevB.97.235301,PhysRevA.68.062308,doi:10.1063/1.1818131}. Because the Hamiltonian of eq.~(\ref{eq:general_hamiltonian}) is time-dependent, we numerically employ a time discretization with step $\Delta t$ to perform the time evolution through the following approximation $U(t, t + \Delta t) \approx \exp \{ -i H(t + \Delta t) \Delta t \}$, which is successively applied from the initial time $t=0$ up to the final time $t=T$. Through the Hamiltonians defined in eqs.~(\ref{eq:Hswap}) and (\ref{eq:Hcnot}), it is possible to find parameters for amplitudes and pulse widths described in eqs.~(\ref{eq:Jswap}) and (\ref{eq:Jcnot}), such that the evolution operator corresponds to the quantum gates. The parameters numerically found are shown in TABLE I and the obtained fidelity exceeds 99.9999\% at the final time, when noise is absent.

\begin{figure}[ht!]
    \includegraphics[width=0.45\textwidth]{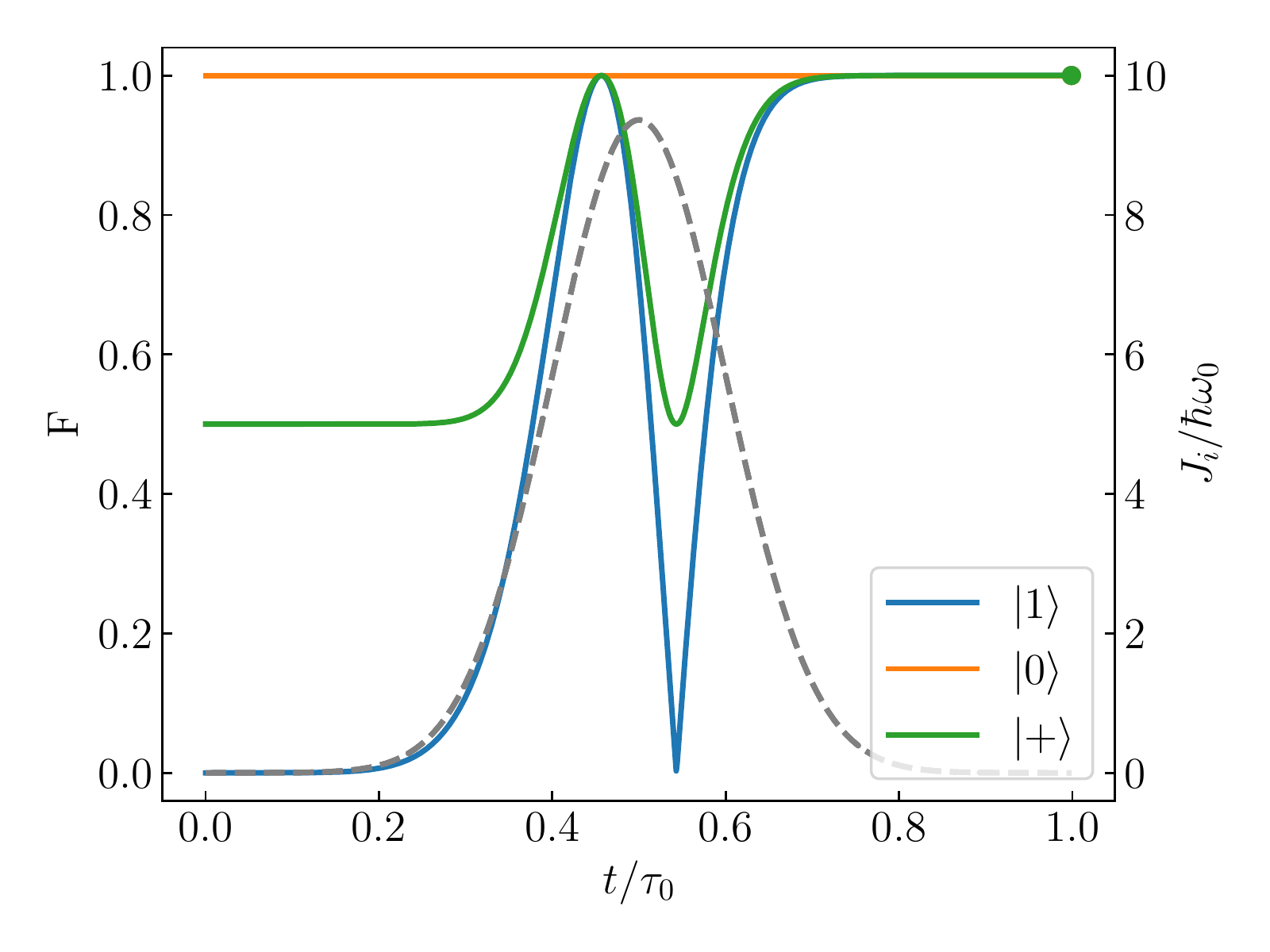}
    \caption{Fidelity $F$ as a function of time $t/\tau_0$ (left axis) for a SWAP gate considering the first qubit $\ket{q_1} = \ket{1}$, $\ket{0}$ and $\ket{+}$ and the second qubit $\ket{q_2} = \ket{0}$. Gaussian pulse $J_{\text{SWAP}}$ as a function of time $t/\tau_0$ (right axis). }
    \label{fig:swap_pulse}
\end{figure}

\begin{figure}[ht!]
    \includegraphics[width=0.45\textwidth]{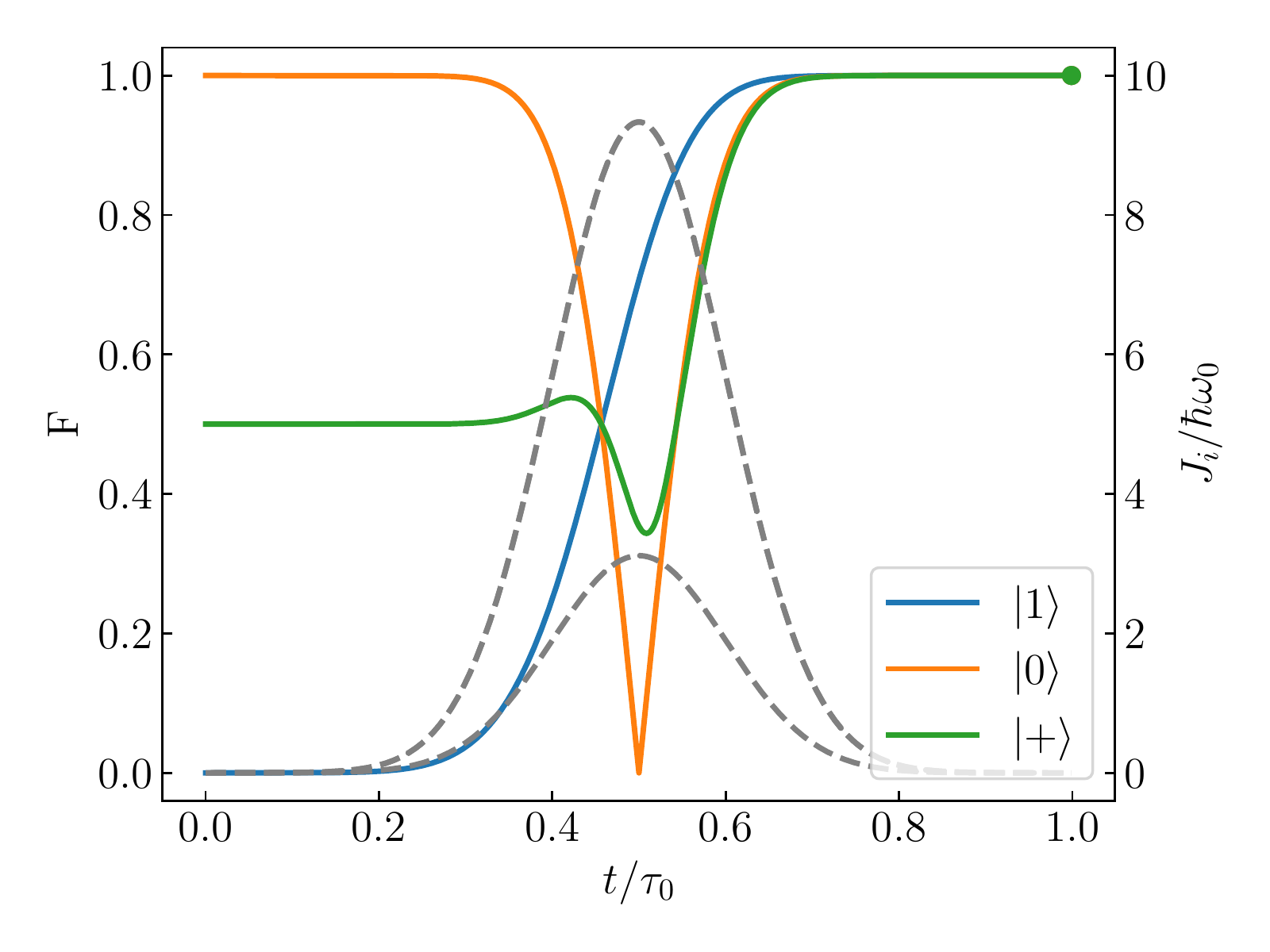}
    \caption{Fidelity $F$ as a function of time $t/\tau_0$ (left axis) for a CNOT gate considering the first qubit $\ket{q_1} = \ket{1}$, $\ket{0}$ and $\ket{+}$ and the second qubit $\ket{q_2} = \ket{0}$. Gaussian pulses $J^1_{\text{CNOT}}$ (higher amplitude) and $J^2_{\text{CNOT}}$ as a function of time $t/\tau_0$ (right axis).}
    \label{fig:cnot_pulse}
\end{figure}

In fig.~(\ref{fig:swap_pulse}), we plot the fidelity for the SWAP gate as a function of time, considering three different initial states of the first qubit $\ket{q_1}=\ket{0}$, $\ket{q_1}=\ket{1}$, and $\ket{q_1}=\ket{+} = (\ket{0}+\ket{1})/\sqrt{2}$, and keeping fixed the second qubit $\ket{q_2} = \ket{0}$. By using the optimized parameters found for the Gaussian pulse, which is shown as a grey dashed curve fig.~(\ref{fig:swap_pulse}), it is possible to see that for all three initial input states the final fidelity approaches one with an error of 10$^{-4}$\% at the final time. Similarly, in fig.~(\ref{fig:cnot_pulse}), we plot the fidelity for the CNOT gate and the respective Gaussian pulses (grey dashed curves) as a function of time. At the final time of evolution, the fidelity exceeds 0.999999 for the same initial states employed in the SWAP case.
\begin{table}[ht!]
\begin{tabular}{|c|c|}
\hline
$A_{\text{SWAP}}$ & 9.36309696 $\hbar\omega_0$   \\ \hline 
$W_{\text{SWAP}}$        & 0.020165$\tau^2_0$ \\ \hline 
$A^1_{\text{CNOT}}$         & 9.33360747$\hbar\omega_0$\\  \hline
$W^1_{\text{CNOT}}$ & 0.02029270$\tau_0^2$ \\ \hline
$A^2_{\text{CNOT}}$          &3.11530553$\hbar\omega_0$          \\ \hline
$W^2_{\text{CNOT}}$  & 0.02023955$\tau_0^2$ \\ \hline
\end{tabular}
\caption{Amplitude and width values for the pulses of SWAP and CNOT gates.}
\label{tab:AL-values}
\end{table}

\subsection{Noise effects}
\begin{figure}
    \centering
    \includegraphics[width=0.45\textwidth]{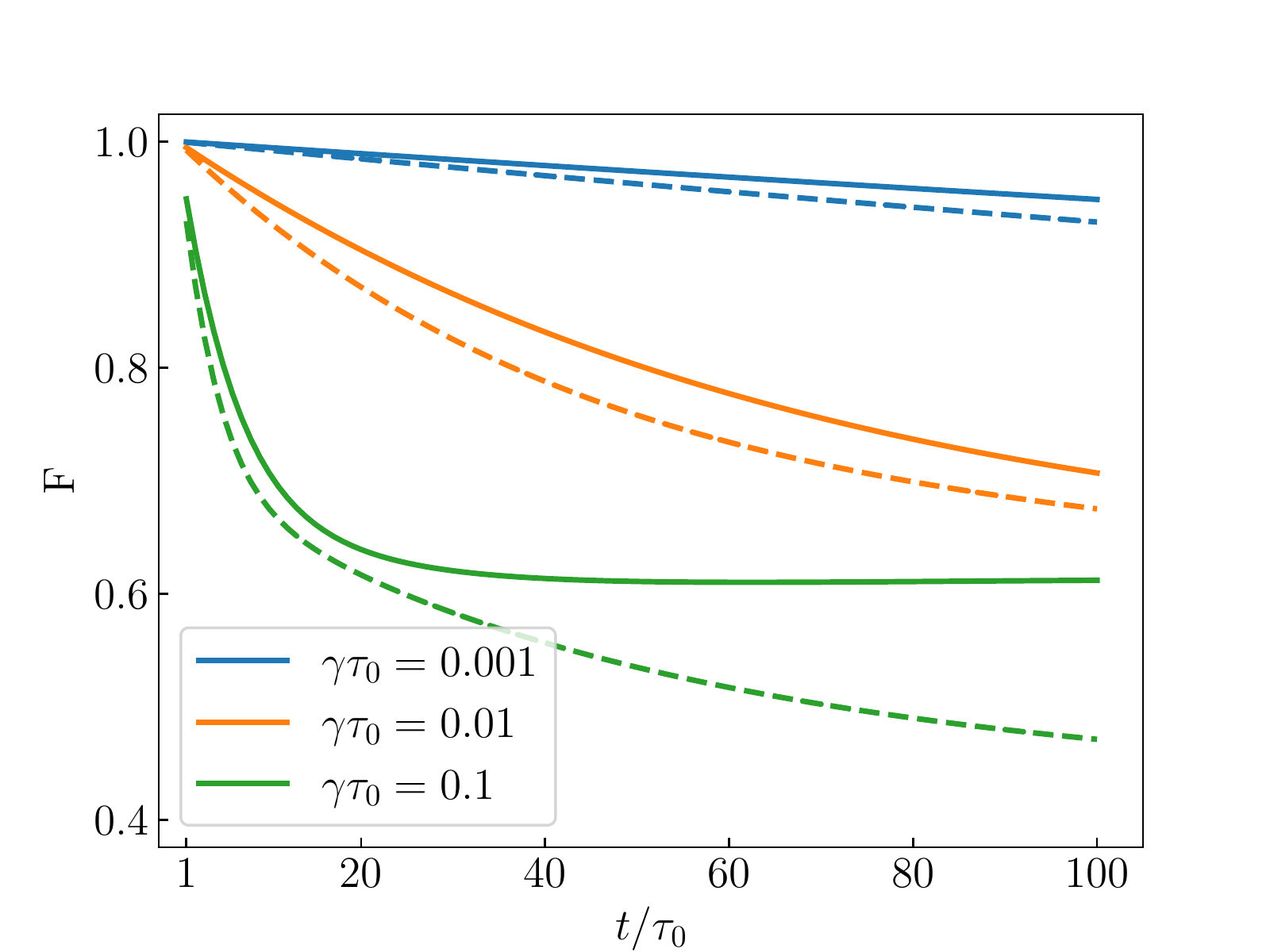}
    \caption{Fidelity $F$ as a function of the final time $t/\tau_0$ for different decay rates $\gamma \tau_0$ = 0.001, 0.01 and 0.1, considering the dephasing type of error. The solid curves correspond to the SWAP gate and the dashed curves correspond to the CNOT gate. The input state is $\ket{q_1} = \ket{+}$ and $\ket{q_2} = \ket{0}$ for both gates.}
    \label{fig:fidelity_by_time_swap_cnot}
\end{figure}

This section aims at understanding the effects of noise during the time evolution of the Gaussian pulses that implement the quantum gates.
By numerically solving eq.~(\ref{eq:lindblad_master_equation}), it is possible to calculate the fidelity at the final time for different decay rates. We also probe these effects by analyzing the final time dependence of the fidelity. To do so, we must rescale the one-pulse duration, center and amplitude such that the fidelity would give one in the absence of noise at the final time. Thus, the parameters in TABLE I must be replaced by $A \rightarrow A/\alpha$ and $W \rightarrow W \alpha^2$, where the one-pulse duration is T=$\alpha\tau_0$ and the center position is $\tau=\alpha\tau_0/2$. In fig.~ (\ref{fig:fidelity_by_time_swap_cnot}), we plot the fidelity $F$ for different final times, considering the dephasing type of noise, which is described by the Lindblad operator $L_{n} = \sqrt{\gamma} \sigma_{n}^{z}$. Also, we consider the implementation of SWAP and CNOT gates over the initial state $\ket{q_1} \ket{q_2}=\ket{+} \ket{0}$, whose fidelity is given by solid and dashed curves, respectively. When the final time is T=$\tau_0$, the fidelity for the SWAP (CNOT) decays to 0.9996 (0.99987), 0.996 (0.9987), and 0.967 (0.987) for $\gamma \tau_0 = 0.001$, 0.001, and 0.1, respectively. As expected, the fidelity decreases as the pulse duration increases, but for $\gamma \tau_0 = 0.1$, the fidelity converges to approximately $F=0.6$ ($F=0.5$) for the SWAP (CNOPT) gate and long pulse duration. This behavior is related to the reaching of the stationary states for both cases.
\begin{figure}
    \includegraphics[width=0.45\textwidth]{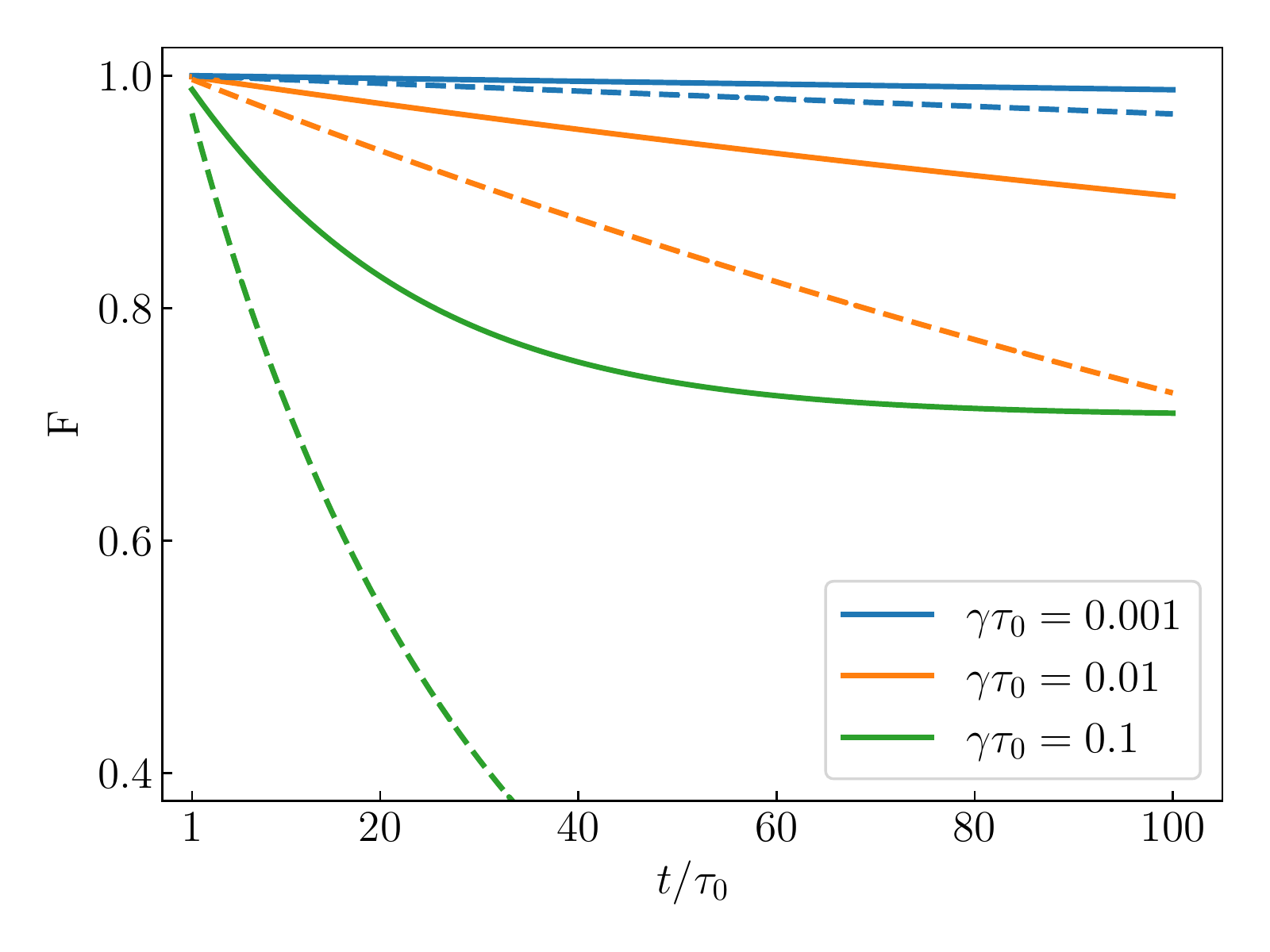}
   \caption{Fidelity $F$ as a function of the final time $t/\tau_0$ for different decay rates $\gamma \tau_0$ = 0.001, 0.01 and 0.1, considering the amplitude-damping type of error. The solid curves correspond to the SWAP gate and the dashed curves correspond to the CNOT gate. The input state is $\ket{q_1} = \ket{+}$ and $\ket{q_2} = \ket{0}$ for both gates.}
    \label{fig:fidelity_by_time_swap_cnot_sigmam}
\end{figure}
Similarly, in fig.~(\ref{fig:fidelity_by_time_swap_cnot_sigmam}) we plot the fidelity $F$ as a function of the final time $T/\tau_0$, but now for the amplitude damping type of noise, where the Lindblad operator is $L_n = \sqrt{\gamma} \sigma_{n}^{-}$, considering the same initial state $\ket{q_1} \ket{q_2}=\ket{+} \ket{0}$. Also, the fidelity decays as the final time increases. However, the SWAP gate experiences a higher robustness when compared to the CNOT gate. For example, the fidelity converges to approximately $F=0.7$ for the SWAP gate, while it goes to zero for the CNOT gate when T=$100\tau_0$ and $\gamma \tau_0 = 0.1$.

\subsection{N-qubit system}
\begin{figure}[ht!]
    \includegraphics[width=0.38\textwidth]{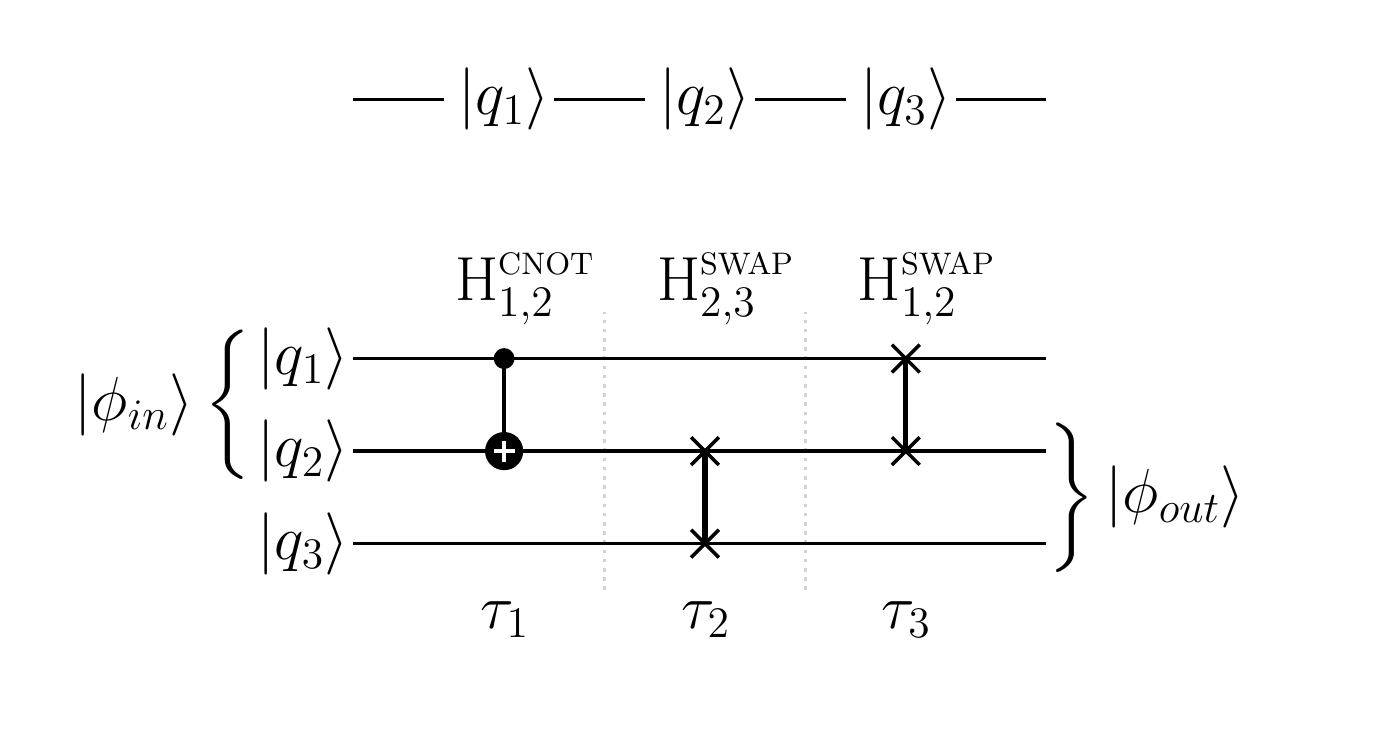}
    \caption{(top panel) Three qubits arranged in a 1D spin chain. (bottom panel) Quantum circuit where a CNOT gate is applied to the first two qubits, $\ket{q_1}$ and $\ket{q_2}$, in the final time $\tau_1$, followed by two SWAP gates applied in the qubits $\ket{q_2}$ and $\ket{q_3}$ in the final time $\tau_2$ and qubits $\ket{q_1}$ and $\ket{q_2}$ in the final time $\tau_3$.}
    \label{fig:3qubits_1line_circuit}
\end{figure}

\begin{figure}[t!]
    \includegraphics[width=0.48\textwidth]{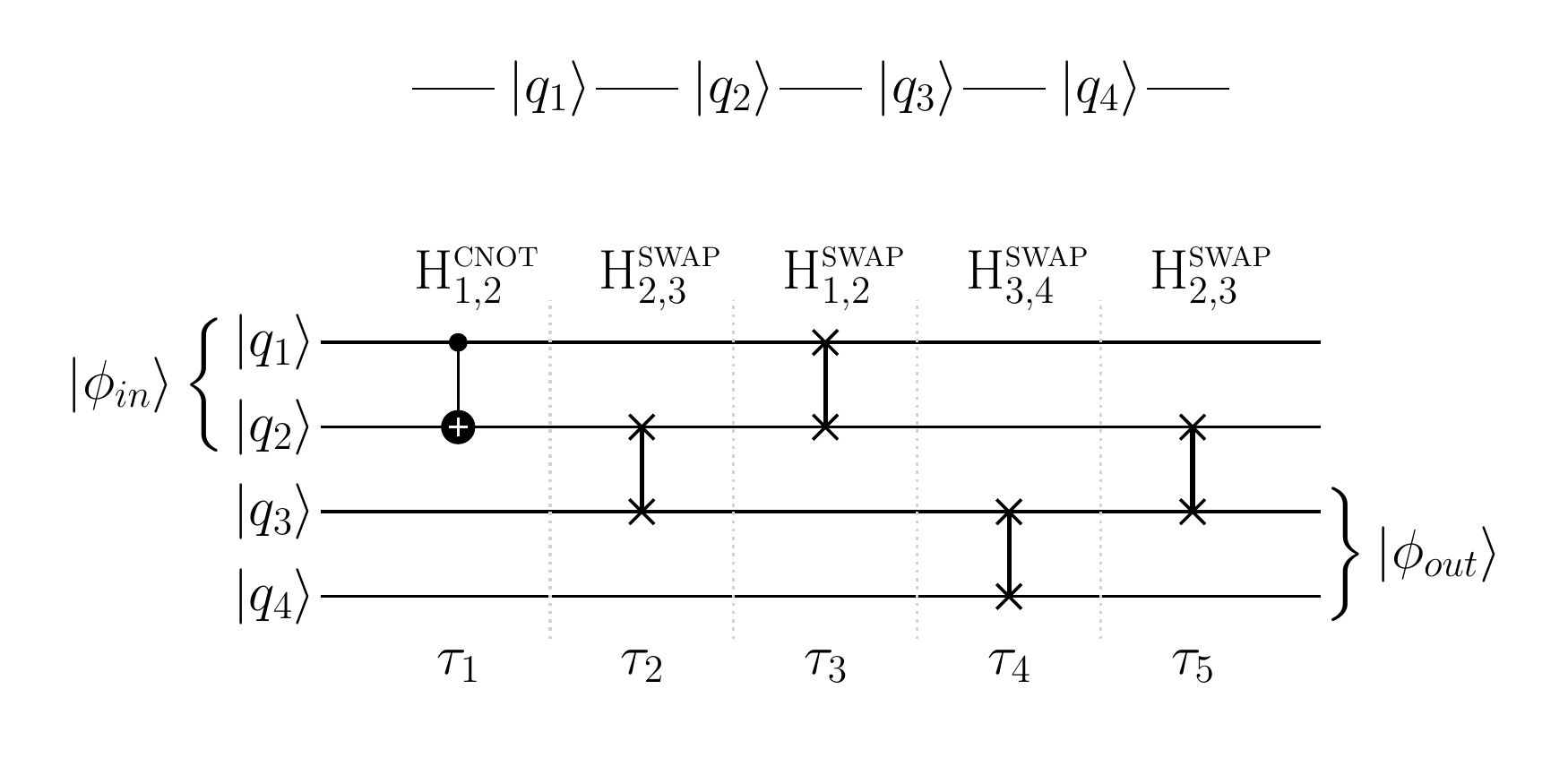}
    \caption{(top panel) Four qubits arranged in a 1D spin chain. (bottom panel) Quantum circuit where a CNOT gate is applied to the first two qubits, $\ket{q_1}$ and $\ket{q_2}$,  followed by SWAP gates that shuttle the quantum information to the last two qubits in the chain. In this case, the k-th pulse occurs at the the final time $\tau_k$, where $k=1,\ldots,5$.}
    \label{fig:4qubits_1line_circuit}
\end{figure}

\begin{figure}[ht!]
    \includegraphics[width=0.45\textwidth]{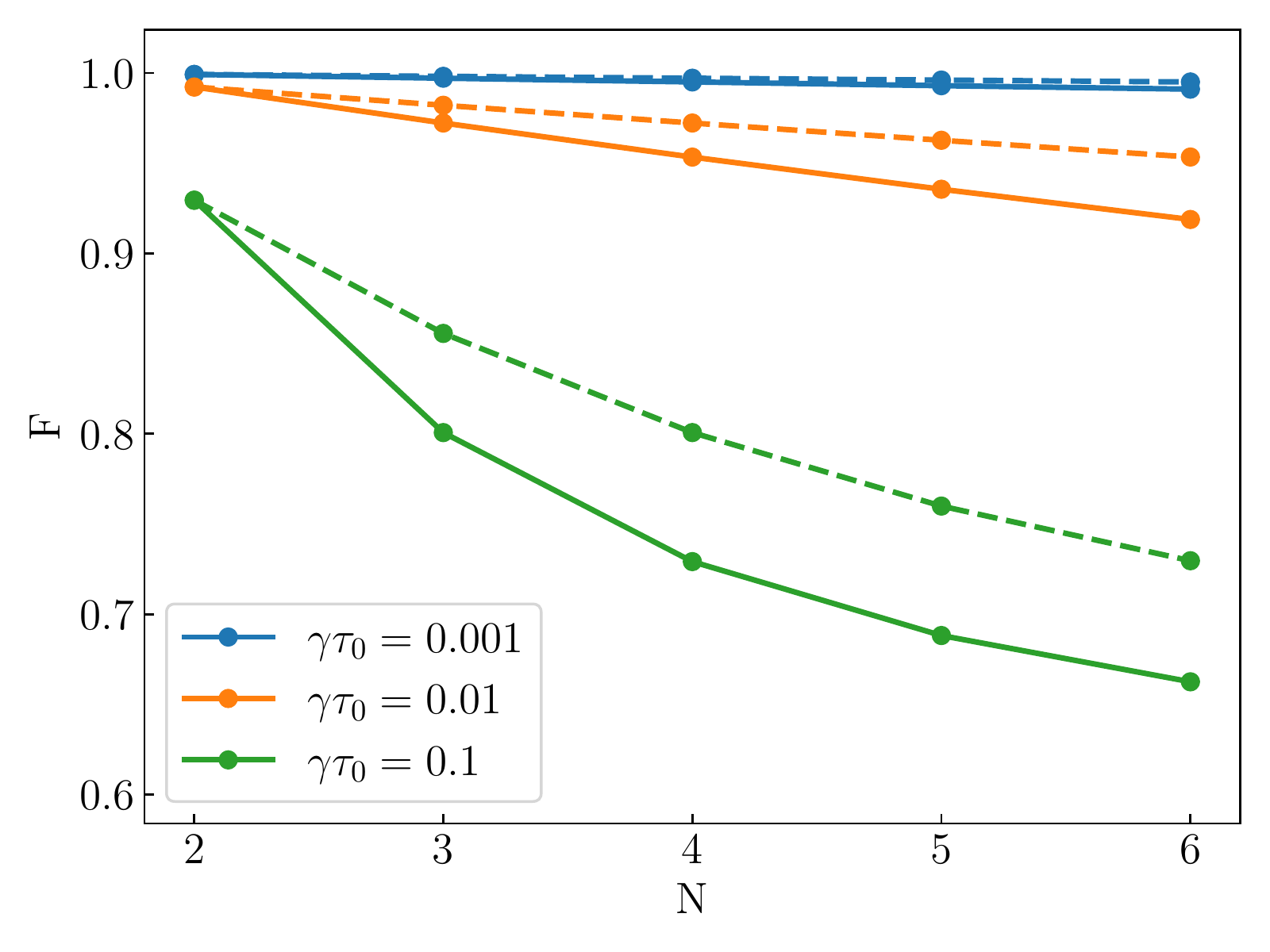}
    \caption{Fidelity $F$ as a function of the number of qubits $N$ for different decay rates $\gamma \tau_0$ = 0.001, 0.01 and 0.1 and the dephasing error. The solid curves correspond to a initial CNOT gate followed by $2(N-2)$ SWAP gates and the dashed curves correspond to a $2(N-2)$ SWAP gates followed by a final CNOT gate considering the qubits in a 1D spin chain. The input state is given by $\ket{q_{1}} = \ket{+}$ with all other qubits $\ket{q_{i}} = \ket{0}$ for $1<i\leq N$.}
    \label{fig:fidelity_by_qubits_1line}
\end{figure}

Here, we explore the connection between different qubits through the implementation of a sequence of gates. The general state for the N-qubit system can be written as $\ket{\psi} = \ket{q_1} \ket{q_2} \dots \ket{q_{N-1}} \ket{q_N}$. We use the first two qubits $\ket{q_1} \ket{q_2}$ as the input state and all other qubits $\ket{q_3} \dots \ket{q_{N-2}}$ as auxiliary qubits. Therefore, the gate implementation will result in a output state that will be at the end of the chain $\ket{q_{N-1}} \ket{q_N}$. When the dynamics is unitary, we can write $\ket{\phi_{in}}$ as the input state and $\ket{\phi_{out}}$ as the output state, which are explicitly given by
\begin{equation}
    \ket{\psi_{in}} = \ket{\phi_{in}} \ket{\phi_{0}^{N-2}},\label{eq.phi_in}
\end{equation}
and
\begin{equation}
    \ket{\psi_{out}} = \ket{\phi_{0}^{N-2}} \ket{\phi_{out}},\label{eq.phi_out}
\end{equation}
where $\ket{\phi_{0}^{N-2}} = \ket{0} \ket{0} \dots \ket{0} \ket{0}$. First, we spatially arrange the qubits in a 1D spin chain, which is a possible design of QDs~\cite{PhysRevX.8.021058}. A two qubit gate is implemented over the state $\ket{\phi_{in}}$, followed by a sequence of SWAP gates that propagates the information from one side to the other of the 1D chain of qubits. To perform this sequence of gates, we use the final time for each pulse as $T=\tau_0$ and the center of the $n$-th pulse obeys the following relation: $\tau_n/\tau_0 = n - 1/2$. To exemplify this arrangement of qubits, we consider three qubits in a 1D spin chain, as represented by the top panel of fig.~(\ref{fig:3qubits_1line_circuit}). In this situation, a CNOT gate is implemented between qubits $\ket{q_1}$ and $\ket{q_2}$ at the final time $\tau_1$ and this information is propagated forward by using a SWAP gate between qubits $\ket{q_2}$ and $\ket{q_3}$, that occurs at the final time $\tau_2$, as depicted in the bottom panel of fig.~(\ref{fig:3qubits_1line_circuit}). Finally, a SWAP gate between qubits $\ket{q_2}$ and $\ket{q_3}$ at the final time $\tau_3$, yields the two-qubit output state mapped onto qubits $\ket{q_2} \ket{q_3}$. If one more qubit is added to the three qubit system, as depicted in fig.~(\ref{fig:4qubits_1line_circuit}), it is necessary to include two more SWAP gates between qubits $\ket{q_3}$ and $\ket{q_4}$ and $\ket{q_2}$ and $\ket{q_3}$ in order to propagate the information to the end of the chain. In conclusion, for $N$ qubits arranged in 1D spin chain, $2(N-2)$ SWAP gates are needed to forward the information from an initial state to the output state $\ket{\phi_{out}} = \ket{q_{N-1}} \ket{q_N}$, when the dynamics is unitary. Of course, when noise is taken into account, eqs.~(\ref{eq.phi_in}) and (\ref{eq.phi_out}) must be replaced by the respective density matrix for N-qubits $\rho^N_{in}$ and $\rho^N_{out}$. The fidelity is evaluated by tracing out all auxiliary qubits, thus $\rho_{out}=\text{ Tr}_{[1,\ldots,N-2]}\{\rho^N_{out}\}$ in eq.~(\ref{eq:fid}).
In fig.~(\ref{fig:fidelity_by_qubits_1line}), we plot the fidelity $F$ as a function of the number of qubits $N$ for different decay rates considering the dephasing type of noise. The qubits are arranged in a 1D spin chain (see fig.~\ref{fig:3qubits_1line_circuit}) and the initial state is $\ket{q_{1}} = \ket{+}$ and $\ket{q_{i}} = \ket{0}$ for $1<i\leq N$. We also analyze the cases where a CNOT gate comes first and it is followed by $2(N-2)$ SWAP gates and the opposite order, {\it i. e.}, $2(N-2)$ SWAP gates followed by a CNOT gate. For both cases, the total time $T$ is proportional to the number of gates and consequently to the number of qubits. When the dynamics is unitary, the fidelity always approaches one (results not shown here). On the other hand, the fidelity decreases with the increasing of the number of qubits for $\gamma \tau_0 = 0.001$, as shown in fig.~(\ref{fig:fidelity_by_time_swap_cnot}). As expected, the fidelity decay is stronger for $\gamma \tau_0 = 0.01$ and $\gamma \tau_0 = 0.1$. Moreover, the fidelity is different if the application of a CNOT gate occurs at the beginning (solid curves) or at the end (dashed curves) of the N-qubits 1D chain. This result is interesting because it shows that order of gates plays an important role when the information is transmitted along the chain, when noise is present.

\begin{figure}[ht!]
    \includegraphics[width=0.35\textwidth]{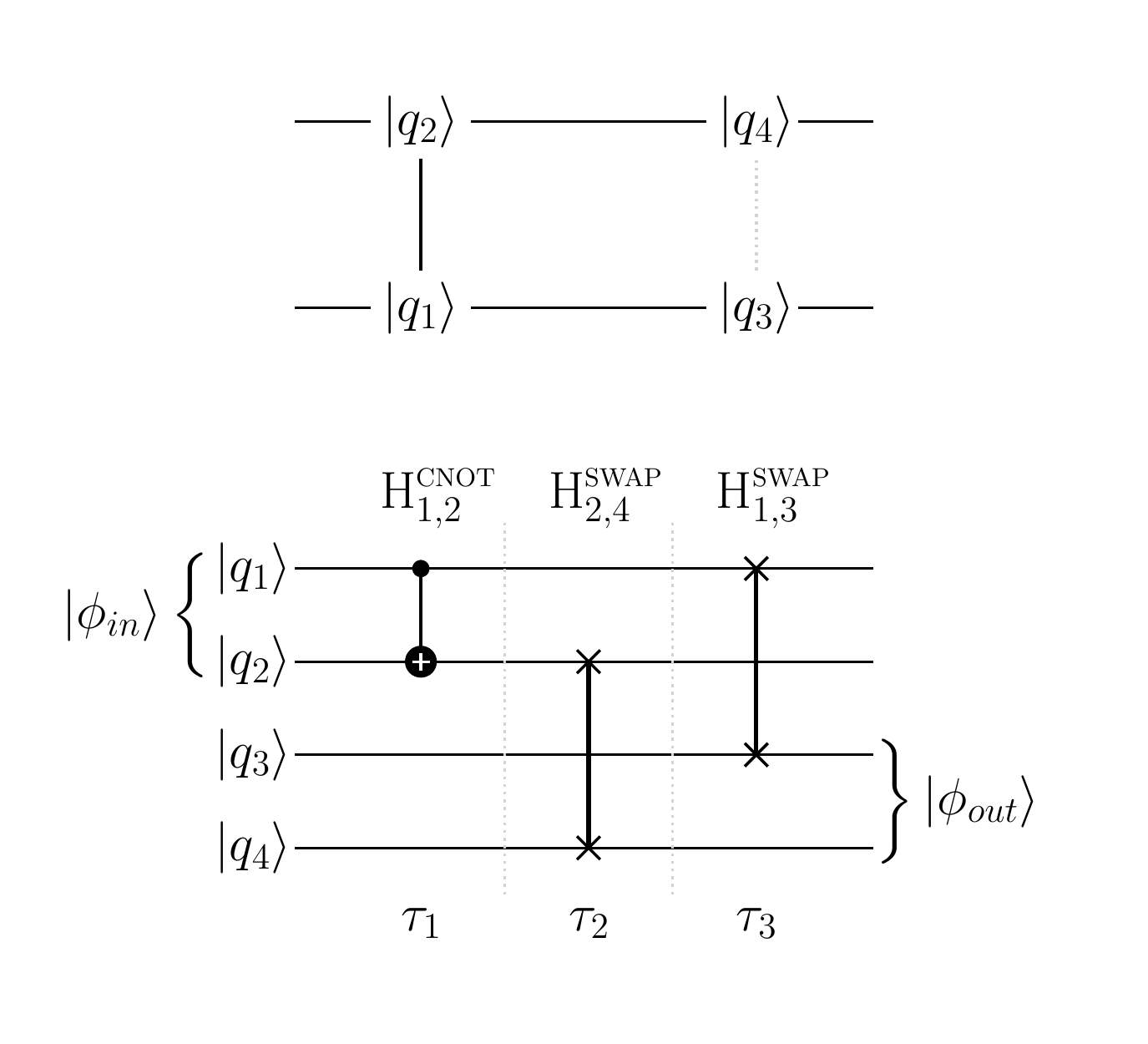}
    \caption{(top panel) Four qubits arranged in a 2D square spin chain. (bottom panel) The quantum circuit where a CNOT gate is applied between the qubits $\ket{q_1}$ and $\ket{q_2}$ in time $\tau_1$, followed by a SWAP gate applied between the qubits $\ket{q_3}$ and $\ket{q_4}$ in the final time $\tau_2$. The last SWAP gate is applied between qubits $\ket{q_1}$ and $\ket{q_3}$ in the final time $\tau_3$.}
    \label{fig:4qubits_2lines_circuit}
\end{figure}

\begin{figure}[ht!]
    \includegraphics[width=0.45\textwidth]{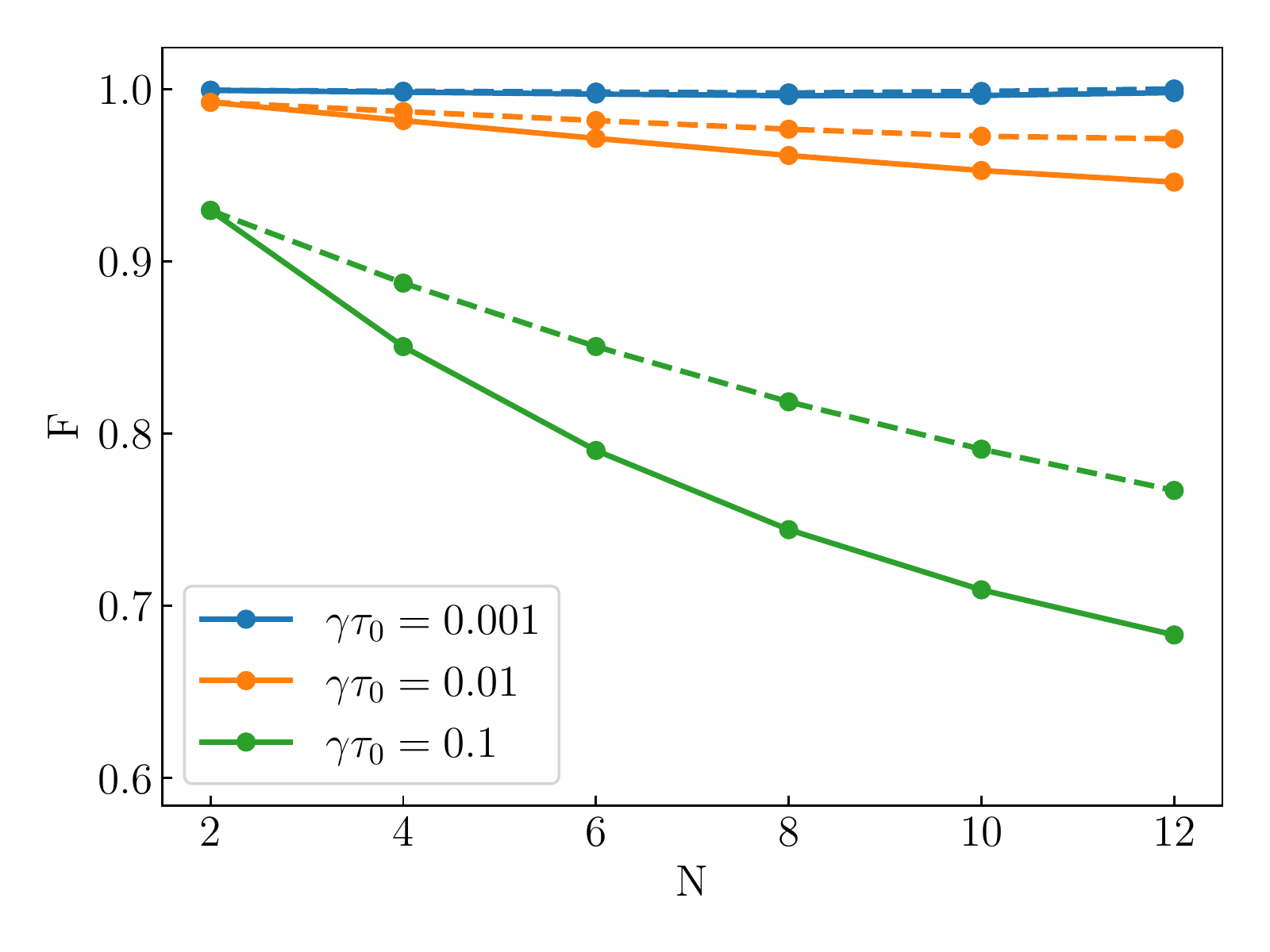}
    \caption{Fidelity $F$ as function of the number of qubits $N$ for different decay rates $\gamma \tau_0$ = 0.001, 0.01 and 0.1, considering the dephasing error. The solid curves correspond to an initial CNOT gate followed by $N/2$ SWAP gates and the dashed curves correspond to $N/2$ SWAP gates followed by a final CNOT gate, considering a 2D square spin chain of qubits. The input state is given by $\ket{q_{1}} = \ket{+}$ with all other qubits $\ket{q_{i}} = \ket{0}$ for $1<i\leq N$.}
    \label{fig:fidelity_by_qubits_2lines}
\end{figure}

For $N$ even, the N-qubit system can also be spatially arranged in a 2D square spin chain of qubits. For example, considering four qubits as depicted in the fig.~(\ref{fig:4qubits_2lines_circuit}), we have the input state $\ket{\phi_{in}} = \ket{q_1} \ket{q_2}$, about which a CNOT gate is applied at the final time $\tau_1$. Two SWAP gates are applied between qubits $\ket{q_2}$ and $\ket{q_4}$ at time $\tau_2$ and between qubits $\ket{q_1}$ and $\ket{q_3}$ at time $\tau_3$. Because these SWAP gates occur between different qubits, they can be consecutively implemented; thus, we can have $\tau_3 = \tau_2$. The two-qubit output are provided by $\ket{\phi_{out}} = \ket{q_3} \ket{q_4}$. For a general case of a 2D square spin chain of $N$ qubits, $N/2$ SWAP gates are needed to forward the information to $\ket{\phi_{out}} = \ket{q_{N-1}} \ket{q_N}$. Therefore, the 2D square spin chain configuration of qubits requires a smaller number of gates to propagate the information and consequently a smaller total time. 
In fig.~(\ref{fig:fidelity_by_qubits_2lines}), we plot the fidelity $F$ as a function of the number of qubits $N$ for different decay rates and the dephasing type of noise. The qubits are arranged in the 2D square spin chain and the initial state is also $\ket{q_{1}} = \ket{+}$ and $\ket{q_{i}} = \ket{0}$ for $1<i\leq N$. Likewise, we consider the situation where a CNOT gate is followed by $N/2$ SWAP gates (solid curves) and vice-versa (dashed curves). Because the SWAP gates are independent in pairs, we can set the time evolution for each pair of pulse respecting the following relation $\tau_{k+1} = \tau_{k}$. Therefore, the total time for propagating the information through the SWAP gates is proportional to $N/2$. Because of this fact, the fidelity for a decay rate $\gamma \tau_0 = 0.1$ for 12 qubits is equivalent to 6 qubits in a 1D spin chain system. This result demonstrates that the 2D arrangement is preferable over the 1D spin chain.
\begin{figure}[ht!]
    \includegraphics[width=0.45\textwidth]{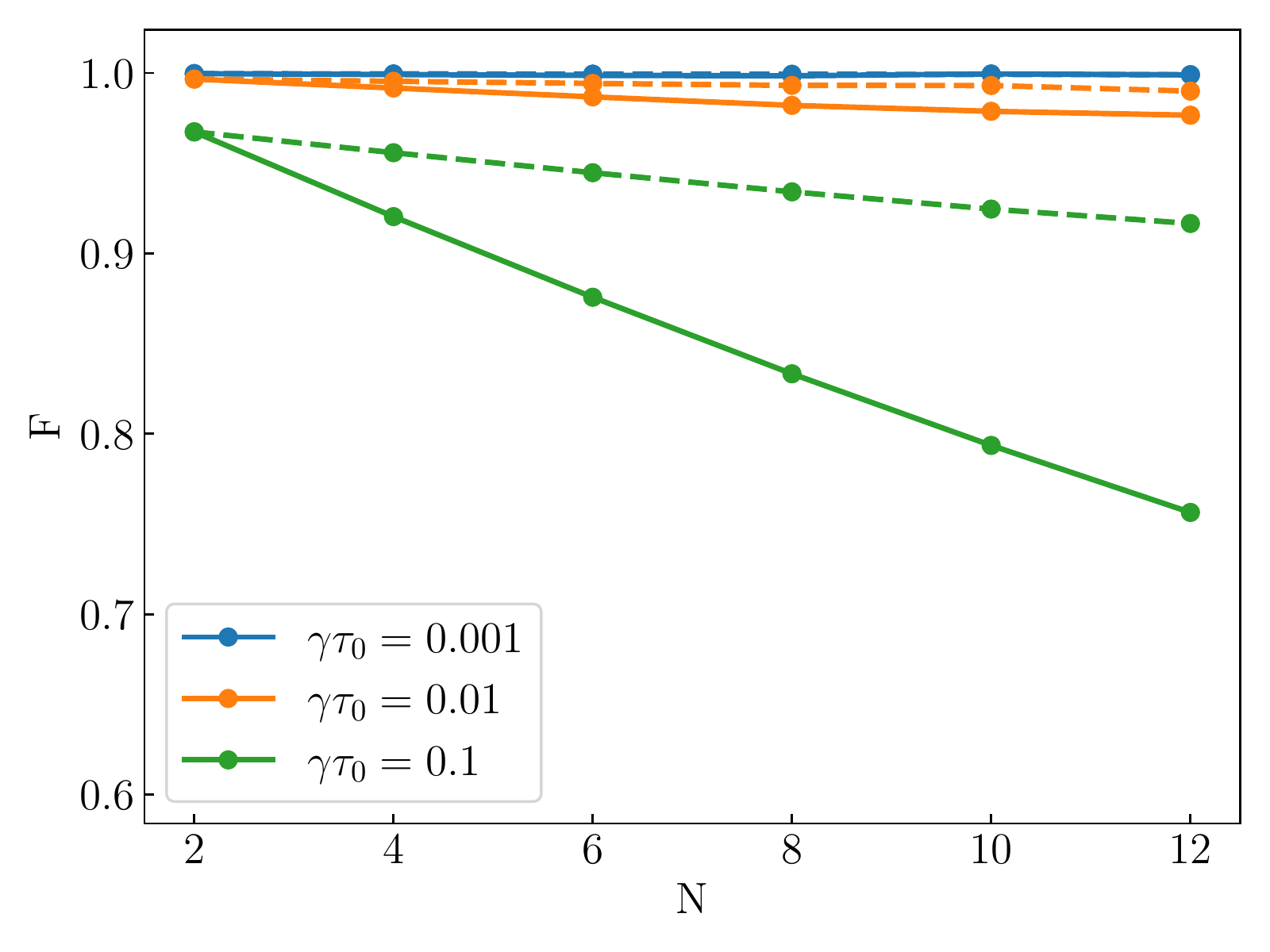}
    \caption{Fidelity $F$ as function of the number of qubits $N$ for different decay rates $\gamma \tau_0$ = 0.001, 0.01 and 0.1, considering the amplitude-damping error. The solid curves correspond to an initial CNOT gate followed by $N/2$ SWAP gates and the dashed curves correspond to $N/2$ SWAP gates followed by a final CNOT gate, considering a 2D square spin chain of qubits. The input state is given by $\ket{q_{1}} = \ket{+}$ with all other qubits $\ket{q_{i}} = \ket{0}$ for $1<i\leq N$.}
    \label{fig:fidelity_by_qubits_2lines_sigmam}
\end{figure}
Similarly, in fig. (\ref{fig:fidelity_by_qubits_2lines_sigmam}), we plot the fidelity as a function of the number of qubits using the 2D square spin chain for the amplitude damping type of noise and same initial state. For small values of the decay rate, $\gamma \tau_0 = 0.001$ and $0.01$, the fidelity is more robust against noise and the fidelity diminishes less than 5\% for $N=12$. For $\gamma \tau_0 = 0.1$, the fidelity presents a linear behavior and the lowest value is approximately $F=0.76$. By the results shown in figs.~(\ref{fig:fidelity_by_qubits_2lines}) and (\ref{fig:fidelity_by_qubits_2lines_sigmam}), we verify that the fidelity is lower when the CNOT gate is applied at the final of the chain for both types of errors and initial state $\ket{q_{1}} = \ket{+}$ and $\ket{q_{i}} = \ket{0}$  for $1<i\leq N$.

\subsection{Initial States}
\begin{figure}[h]
    \includegraphics[width=0.45\textwidth]{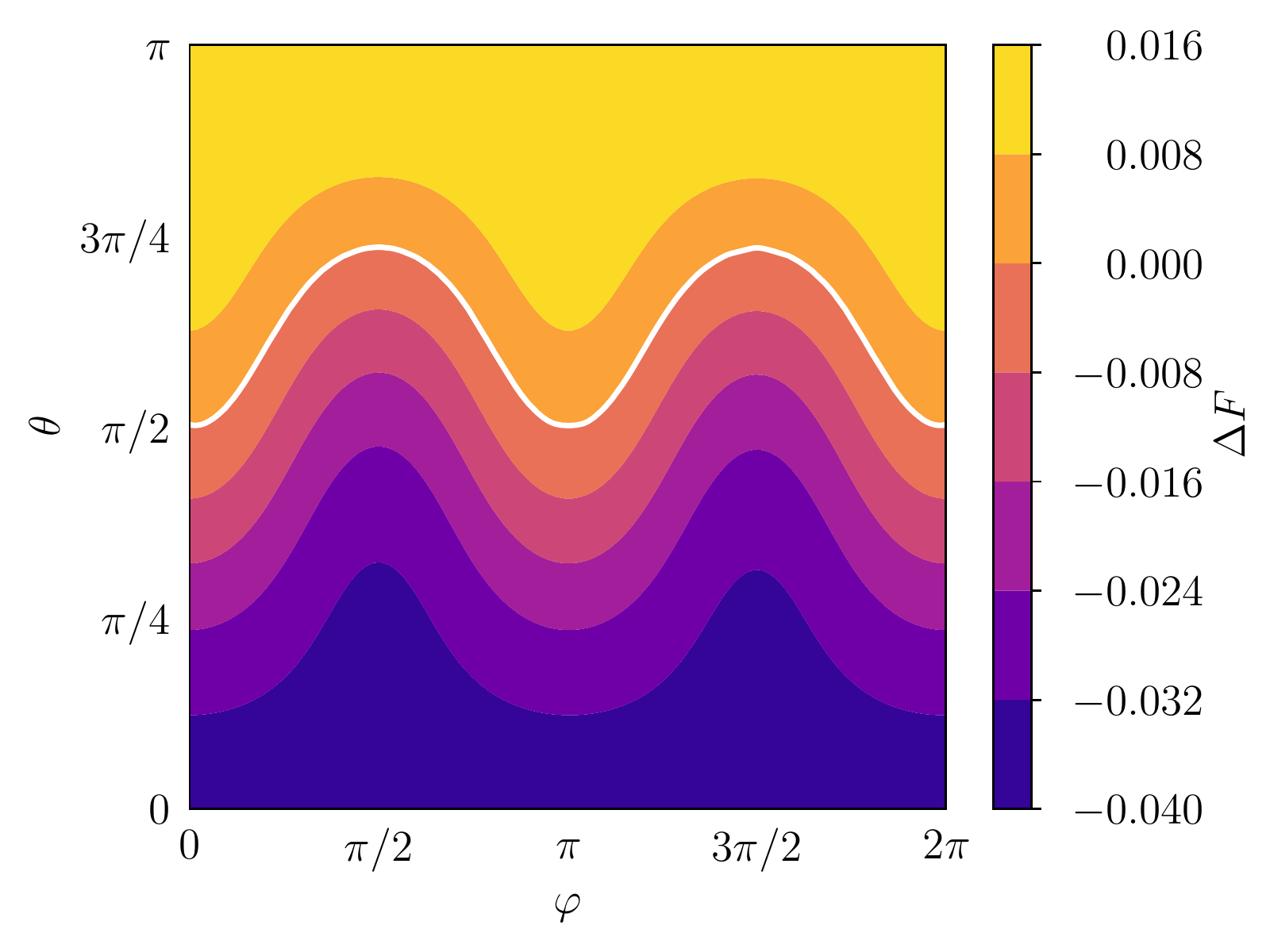}
    \caption{Fidelity difference considering a CNOT followed by two SWAP gates and vice-versa as a function of $\theta$ and $\varphi$, which are related to the initial state $\ket{\phi_{in}(\theta, \varphi)} = \ket{+} \ket{q_2(\theta, \varphi)}$ for decay rate $\gamma \tau_0 = 0.1$ and the amplitude-damping error. In this case,  four qubits are spatially arranged in a 2D square spin chain. The white line represents the values where $\Delta F = 0$.}
    \label{fig:fidelity_difference_q2theta_sm}
\end{figure}

\begin{figure}
    \includegraphics[width=0.45\textwidth]{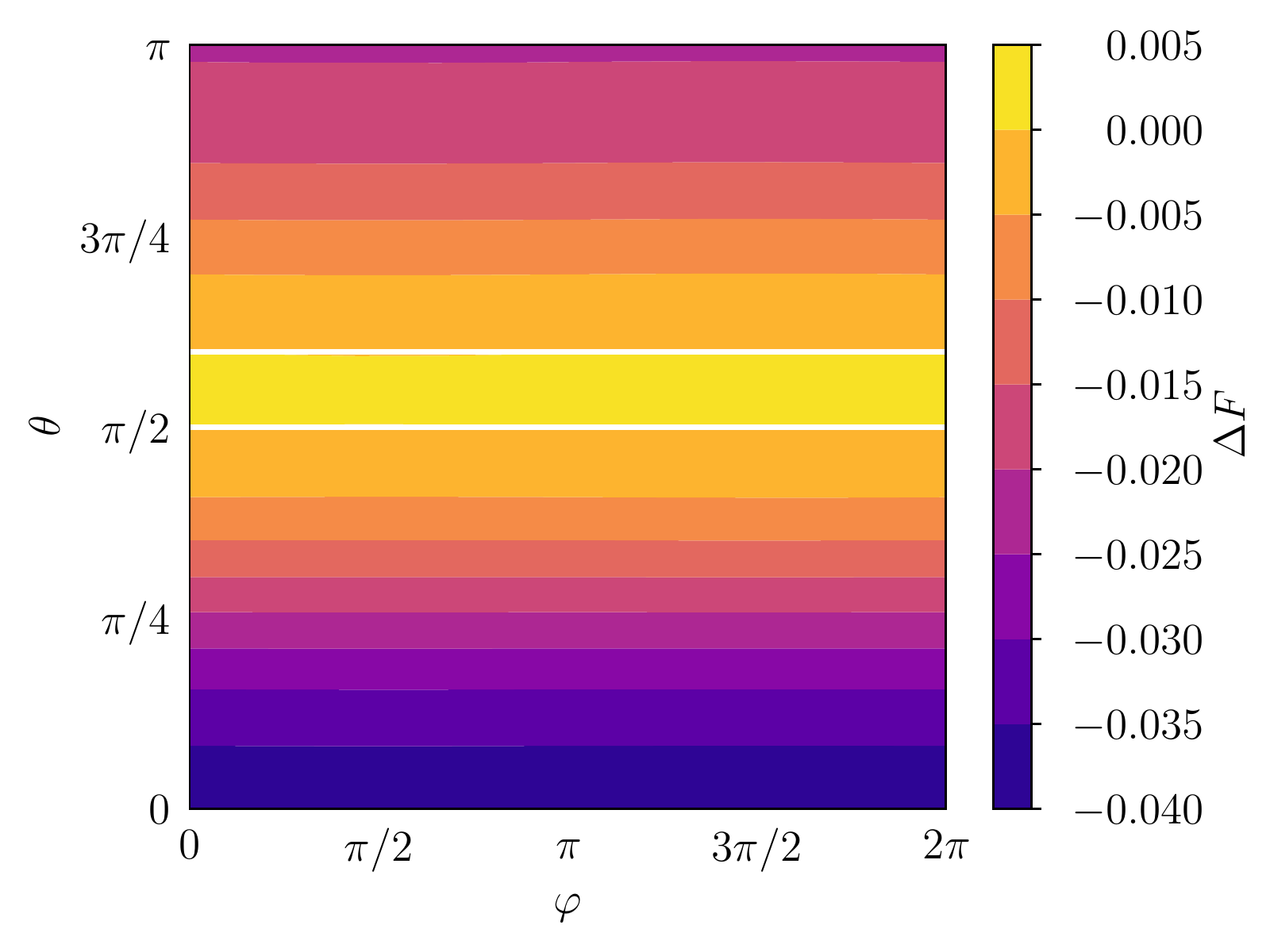}
    \caption{Fidelity difference considering a CNOT followed by two SWAP gates and vice-versa as a function of $\theta$ and $\varphi$, which are related to the initial state $\ket{\phi_{in}(\theta, \varphi)} = \ket{+} \ket{q_2(\theta, \varphi)}$ for decay rate $\gamma \tau_0 = 0.1$ and the dephasing error. In this case, four qubits are spatially arranged in a 2D square spin chain. The white line represents the values where $\Delta F = 0$.}
    \label{fig:fidelity_difference_q2theta_sz}
\end{figure}

In this section, we intend to explore the initial state influence on the fidelity when the order of quantum gates is distinct and noise is taken into account.
To accomplish such a task, we choose the initial input state as $\ket{\phi_{in}(\theta, \varphi)} = \ket{+} \ket{q_2(\theta, \varphi)}$, where
\begin{equation}
    \ket{q_2(\theta, \varphi)} = \sin(\theta/2)\ket{0} + e^{i \varphi} \cos(\theta/2) \ket{1},
\end{equation}
with $\theta \in [0, \pi]$ and $\varphi \in [0, 2\pi]$.
To be able to evaluate the fidelity for several initial states, we choose a 2D square spin chain composed of 4-qubits, as depicted in fig.~(\ref{fig:4qubits_2lines_circuit}).
Furthermore, we calculate the difference of the fidelity evaluated for a CNOT gate followed by two SWAP gates and vice-versa; thus, $\Delta F = F_{\text{1CNOT--2SWAP}} - F_{\text{2SWAP--1CNOT}}$.
 In fig.~(\ref{fig:fidelity_difference_q2theta_sm}) we plot $\Delta F$ as a function of $\theta$ and $\varphi$ in a colormap for decay rate $\gamma \tau_0 = 0.1$ considering the amplitude-damping noise. One can see that the fidelity variation $\Delta F$ is predominantly less than zero, which implies that the 2SWAP-1CNOT order provides a higher fidelity than the one obtained for the 1CNOT-2SWAP order. The solid white curve indicates the angles where $\Delta F = 0$, which can be approximated by the following relation $\theta(\varphi) \approx a\cos{(2\varphi)} + b$.
In fig.~(\ref{fig:fidelity_difference_q2theta_sz}), we plot $\Delta F$ as a function of $\theta$ and $\varphi$, considering the dephasing type of noise. One can see in fig.~(\ref{fig:fidelity_difference_q2theta_sz}) that the fidelity variation depends only on $\theta$. Also, the solid withe curves indicate where $\Delta F = 0$ and define a small region between approximately $\theta = \pi/2$ and $3\pi/5$ where the value of the difference $\Delta F$ is positive. In this case, the order 2SWAP-1CNOT provides a higher fidelity for the majority of initial states when compared to the order 1CNOT-2SWAP. Although the difference between the orders of gates implementation are small in both considered cases, it might be much bigger when a large number of qubits are taken into account.

\section{Conclusions}
In this paper, we propose a scheme to connect long distance QD qubits by using a spin chain with nearest-neighbors interaction considering a Heisenberg type Hamiltonian with time dependent pulses. We employ this approach to obtain the SWAP and the CNOT gates, although other gates could be tested. When the system is free from decoherence, we achieve a fidelity of 0.999999 for both type of gates. We also consider the most important types of noise, dephasing and amplitude-damping, by using the master equation approach. Moreover, we propose two different spatial arrangements for the physical qubits, the 1D chain and the 2D square chain. The 2D chain provides a better performance than the 1D chain, because we can apply pairs of gates at the same time. Furthermore, we investigate the role of the order of gates implementation when decoherence is taken into account, thus, we simulate the action of a CNOT gate followed by a sequence of SWAP gates that transport the quantum information and vice-versa. We found that the order of application of these gates is relevant and depends on the initial state of the system. This difference of order of gates can be very large when the system is composed of a large number of qubits, which is the case in a quantum computer.

\section*{Acknowledgments}
We are grateful for financial support by the Brazilian Agencies FAPESP, CNPq and CAPES. 
LKC acknowledges support from S\~ao Paulo Research Foundation, FAPESP (grant 2019/09624-3) and from National Council for Scientific and Technological Development, CNPq (grant 311450/2019-9). 

\bibliography{ref.bib}
\end{document}